\documentstyle[11pt,aaspp4]{article}

\begin{document}     
\baselineskip 0.202in

\newcommand{\squig}{$\sim$}
\newcommand{\squigleq}{\mbox{$^{<}\mskip-10.5mu_\sim$}}
\newcommand{\squiggeq}{\mbox{$^{>}\mskip-10.5mu_\sim$}}
\newcommand{\squiggeqmm}{\mbox{$^{>}\mskip-10.5mu_\sim$}}
\newcommand{\decsec}[2]{$#1\mbox{$''\mskip-7.6mu.\,$}#2$}
\newcommand{\decsecmm}[2]{#1\mbox{$''\mskip-7.6mu.\,$}#2}
\newcommand{\decdeg}[2]{$#1\mbox{$^\circ\mskip-6.6mu.\,$}#2$}
\newcommand{\decdegmm}[2]{#1\mbox{$^\circ\mskip-6.6mu.\,$}#2}
\newcommand{\decsectim}[2]{$#1\mbox{$^{\rm s}\mskip-6.3mu.\,$}#2$}
\newcommand{\decmin}[2]{$#1\mbox{$'\mskip-5.6mu.$}#2$}
\newcommand{\asecbyasec}[2]{#1$''\times$#2$''$}
\newcommand{\aminbyamin}[2]{#1$'\times$#2$'$}

\title{A Search for the Optical Counterpart of the Luminous X-ray
Source in NGC\,6652\,\footnote{\ Based on observations with the NASA/ESA
Hubble Space Telescope, obtained at the Space Telescope Science
Institute, which is operated by the Association of Universities for
Research in Astronomy, Inc., under NASA contract NAS5-26555.}
}
\author{Eric W. Deutsch, Bruce Margon, and Scott F. Anderson}
\affil{Department of Astronomy, 
       University of Washington, Box 351580,
       Seattle, WA 98195-1580\\
       deutsch@astro.washington.edu; margon@astro.washington.edu;
       anderson@astro.washington.edu}

%\begin{center}
%Submitted for publication in the Astronomical Journal\\
%To appear in volume 493, 1998 February 1\\
%{\it received 1998 April 8}
%\end{center}

\begin{center}
Accepted for publication in The Astronomical Journal\\
To appear in volume 116, September 1998\\
{\it received 1998 April 8; accepted 1998 May 19}
\end{center}

%==============================================================================
\begin{abstract}

We examine images of the field of X1832--330, the luminous (${\rm
L_X\sim10^{36}\ erg\ s^{-1}}$) X-ray burst source near the center of
the globular cluster NGC\,6652, in order to identify the optical
counterpart for further study.  U and B ground-based images allow us to
set a limit $M_{B_0}\squiggeqmm3.5$ for the counterpart at the time of
those observations, provided that the color is $(U-B)_0\sim-1$, similar to
the sources known in other clusters.  Archival {\it Hubble Space
Telescope} observations survey most but not all of the $1\sigma$ X-ray
error circle, and allow us to set limits $M_{B_0}>5.9$ and
$M_{B_0}>5.2$ in the WF/PC and WFPC2 regions, respectively.

In the WF/PC images we do weakly detect a faint object with UV-excess,
but it is located \decsec{11}{7} from the {\it ROSAT} X-ray position.
This considerable ($2.3\sigma$) discrepancy in position suggests that
this candidate be treated with caution, but it remains the only
reasonable one advanced thus far.  We measure for this star
$m_{439}=20.2\pm0.2$, $(m_{336}-m_{439})=-0.5\pm0.2$, and estimate
$M_{B_0}=5.5$, $(U-B)_0=-0.9$, similar to other known optical
counterparts.  If this candidate is not the identification, our limits
imply that the true counterpart, not yet identified, is probably the
optically-faintest cluster source yet known, or alternatively that it
did not show significant UV excess at the time of these observations.
Finally, we assess the outlook for the identification of the remaining
luminous globular cluster X-ray sources.

\end{abstract}


%==============================================================================
\clearpage
\section{INTRODUCTION}

While the X-ray properties of the luminous globular cluster X-ray
sources have been studied for over two decades, only in the last five
years has there been significant progress in the study of their optical
counterparts.  During this period the situation has gone from the
existence of only one identification of an unusually optically-luminous
system in M\,15, to confirmed or likely optical counterparts in five
clusters (Deutsch et al. 1998 and references therein).  {\it Hubble
Space Telescope (HST)} has largely been responsible for this advance,
principally due to the extreme crowding in these fields, which limits
the utility of ground-based programs.

Globular cluster X-ray sources are interesting targets for study for
many reasons.  It has been known for over two decades that globular
cluster X-ray sources are overabundant with respect to those in the
field (Katz 1975; Clark 1975); while globular clusters contribute only
a tiny fraction to the total number of stars in the Galaxy, 10\% of the
known low-mass X-ray binaries (LMXBs) are found in globular clusters
(van Paradijs 1995).  It is still not clear whether these systems are
somehow different as a group from those in the field, or instead the
globular cluster environment merely enhances their formation
probability.  Indeed, it has long been suspected that close binaries
may dominate the binding energy in globular clusters, and these exotic
binaries may hold important clues to binary formation and interaction
in these clusters.  The number and properties of clusters containing
LMXBs have been used to test stellar interaction hypotheses (e.g.,
Verbunt \& Hut 1987, Predehl et al. 1991).  In addition, luminosities
and intrinsic colors may be determined far more accurately than for
field sources, as the distances and reddenings to the host clusters can
be readily determined.  Identification and further study of the optical
counterparts of these sources provide many more opportunities to
determine system parameters and unravel the nature of these LMXBs than
can X-ray observations alone.

Here we present results on a search for the optical counterpart, for
which there is as yet no candidate, to X1832--330, the luminous (${\rm
L_X\sim10^{36}\ erg\ s^{-1}}$) X-ray burst source near the center of
the globular cluster NGC\,6652.  This paper extends the preliminary
work presented by Deutsch et al. (1997).

This X-ray source was probably first detected as H1825--331 in the
HEAO-1 survey.  However, as the 2.7 deg$^2$ 90\% confidence error box
contained the cluster, but was very close to the Galactic Center, Hertz
\& Wood (1984) conservatively allowed that the association with
NGC\,6652 was premature.  With the significantly-better spatial
resolution of the {\it ROSAT} All-Sky Survey (RASS), Predehl et al.
(1991) reported a bright X-ray source which was indeed coincident
with NGC\,6652 to 1$'$.  Using reprocessed RASS data, Verbunt et al.
(1995) estimate that the flux was as much as $\sim10\times$ higher
(depending on spectral assumptions) than during the HEAO-1 detection.  Pointed
observations with the {\it ROSAT} PSPC 1.5 yr after the RASS find the
source somewhat brighter still, and show \squig20\% variations on a
time scale of a few hours (Johnston et al. 1996).  X-ray luminosity
estimates inferred from these various observations span the range ${\rm
L_X=10^{35} - 10^{36}\ erg\ s^{-1}}$, all normalized to the 0.5--2.5
keV band and distance adopted below.  This luminosity variation
prompted Verbunt et al. (1995) to label this source as a transient,
although the variability observed thus far seems orders of magnitude
less than that of sources indisputably called transients.  Most
luminous globular cluster X-ray sources are known to be bursters,
indicating that the primaries are neutron stars.  Just recently
X1832--330 joined the ranks of known bursters when two Type I
bursts were reported by in 't Zand et al. (1998).

Ortolani et al. (1994) carried out the first color-magnitude study of
NGC\,6652.  They derive $({\rm m-M})_0=14.85$ (${\rm d=9.3}$ kpc),
E($B-V)=0.10$, and estimate [Fe/H$]\approx-0.9$.  These values are
similar to previous determinations except for the distance, which is
30\% closer.  From the compilation of Trager et al. (1993), we adopt a
core radius $r_c=$\ \decsec{4}{3}.  X1832--330 should be a relatively
easy target for a search for an optical counterpart as NGC\,6652 is
neither heavily reddened nor extremely dense, whereas most of the
clusters which harbor luminous X-ray sources do fall into one or both
of these two categories.  However, as the cluster is near the Galactic
center ($l=\decdegmm{1}{5}$, $b=\decdegmm{-11}{4}$) the contamination
of the field by bulge stars is of some concern.  A brief study of this
issue is presented by Ortolani et al.  Finally we note that in their
color-magnitude diagram, Ortolani et al. indicate a sizable population
of blue stragglers.

%==============================================================================
\section{OBSERVATIONS AND DATA REDUCTION}

\subsection{X-ray Coordinates}

While the initially-reported RASS coordinates had only $1'$ accuracy,
they were refined by Verbunt et al. (1995) to a $1\sigma$ uncertainty
of $20''$; even the latter would make a search for an optical counterpart
difficult.  However, a short pointed {\it ROSAT} HRI observation of
NGC\,6652 obtained on 1994 March 28 is available in the archive.  From
these observations, we measure coordinates
$\alpha(2000)={\rm18^h35^m}$\decsectim{44}{0},
$\delta(2000)={\rm-32^\circ59'}29''$, which agree with the position
given as preliminary by Grindlay (1995).

The uncertainty of these coordinates is almost entirely due to the
uncertain telescope aspect, as the source is well detected.  From a
study by Voges (1998) of the correlation between {\it ROSAT} HRI source
positions and stars in the Tycho Catalog (ESA, 1997), we infer based on
\squig 3500 matches that the $1\sigma$, 90\% confidence, and $2\sigma$
uncertainties are \decsec{5}{3}, \decsec{8}{3}, and \decsec{10}{2},
respectively.  These values are consistent with the analysis of
HRI uncertainties by David et al. (1992).

These coordinates are more uncertain than the {\it Einstein}
HRI positions available for some sources in other clusters.
Unfortunately no X-ray sources appear in the {\it ROSAT} HRI exposure
other than the cluster source, so correction of the X-ray aspect
via association with other optical identifications, e.g. bright
foreground stars, is not possible.  However, the coordinates are
centered \decsec{20}{5} (4.8 $r_c$) from the cluster center where the
crowding is not severe by {\it HST} standards, so the
search of a large error circle is still feasible.

\subsection{Ground-Based Imagery}

To find the precise optical position of the X-ray coordinates, we
create an astrometric solution for a ground-based B-band CCD image
(kindly provided by S. Wachter) by using 21 isolated stars in common
between the CCD image and the Digitized Sky Survey, which is based on
the {\it HST} Guide Star Catalog (GSC) reference frame.  The
uncertainty in this transfer is \decsec{0}{04}.  With this transferred
coordinate system, we place a cross at our optical position for the
X-ray coordinates in Fig. 1.  The systematic uncertainty in the GSC
frame with respect to other astrometric frames for a typical region is
\squig \decsec{0}{5} (Russell et al. 1990).  Overlaid in Fig. 1 are
\decsec{5}{3} ($1\sigma$) and \decsec{10}{2} ($2\sigma$) radius X-ray
error circles.  From this CCD image, obtained in \decsec{1}{3} seeing
conditions, it is clear that the field is quite crowded from the
ground.  As only one color was obtained, these data are not directly
suitable for searching for an optical counterpart.

On 10 June 1997 we obtained brief U (300 s) and B (100 s) exposures of
NGC\,6652 with the Taurus Tunable Filter (TTF) (Bland-Hawthorn \& Jones
1998) in a broadband imaging mode on the Anglo-Australian Telescope (AAT).
Despite the \decsec{3}{5} seeing and cirrus conditions at the time of
the exposures, these frames are deep enough to search part of the range
of expected luminosities.  We subtract the two frames in order to
search for some excess UV flux in the error circles, but find none.  If
we assume that the counterpart has $(U-B)_0\sim-1$, similar to the other
known cluster counterparts, we are able to set a lower limit
$M_{B_0}\squiggeqmm3.5$ within the entire $2\sigma$ X-ray uncertainty
radius, using artificial star tests.  Three of the five currently-known
optical counterparts have luminosities brighter than this (Deutsch et
al. 1998), and therefore could have been detected with these data, had
they been in NGC\,6652.

\subsection{{\it HST} Imagery}

The recent success in discovering optical counterparts to the globular
cluster X-ray sources has largely been due to their considerable UV excess,
$(U-B)_0\sim-1.0$, observed in {\it HST} images.  We therefore take
this same approach here, and search for UV-excess objects in a
color-magnitude diagram of the cluster.  {\it HST} WF/PC and WFPC2
observations of this cluster, taken as part of unrelated programs,
are available in the {\it HST} data archive, and we have extracted
these images for this work.  The {\it HST} fields of view are
overlaid on the ground-based CCD frame in Fig. 1.  The WF/PC
observations were obtained at three different offsets and exposure
times for each filter, while the WFPC2 observations were all at the
same pointing.  It can be seen from the figure that, simply by chance,
the X-ray error circles extend somewhat beyond the boundaries of the
{\it HST} frames.  While these data are clearly not optimal, there is a
reasonably good chance of success in isolating the optical counterpart
to X1832--330.

To place the position of the X-ray coordinates on the {\it HST} images,
we transfer the astrometric solution from the CTIO CCD image discussed
in \S2.2 to these images.  The internal errors in the transfer are
negligible compared to the systematic uncertainties in the GSC frame as
discussed above.

For the WFPC2 images, measured magnitudes are calibrated to the STMAG
system using the photometric zero points from Table 9 ($Z_{STMAG}$) in
Holtzman et al. (1995a).  Aperture corrections are taken from Table
2(a) in Holtzman et al. (1995b).  There has not been any correction
applied for geometric distortions or charge transfer efficiency losses
(Holtzman et al. 1995b) as the small errors introduced by these effects
are not of concern for our purposes here.

The WFPC2 observations used the F218W (500 s, 900 s), F439W (50 s, 2
$\times$ 160 s), and F555W (10 s, 50 s) filters; the latter two closely
resemble Johnson B and V filters, respectively.  There are no objects
detected within 15$''$ of the X-ray position in the F218W exposures
(which do not image the entire error circle), with lower limit
$m_{218}>19.7$.  If we assume that the optical counterpart has
$(m_{218}-m_{439})_0=-1.4$, similar to the optical counterpart in
NGC\,6441 (Deutsch et al. 1998), we infer a lower limit of
$m_{439}>20.5$ or $M_{B_0}>5.2$ in the region of the error circles
subtended by the WFPC2 exposures.  This value may be compared with
$M_{B_0}=5.6$ for the least luminous globular cluster counterpart, Star
A in NGC 1851 (Deutsch et al. 1998).

As the optical counterparts in other clusters have been strongly UV
excess with unremarkable $B$, $V$ colors, the F439W and F555W images are
not suitable for isolating a counterpart based on color alone.
However, observed variability can be used to identify a counterpart, as
optical/UV variability has been detected for all of the known
counterparts for which a suitable study has been performed.
Unfortunately, these short consecutive snapshots are unlikely to allow
detection of anything but extremely-short timescale, high-amplitude
variability.  The images have been examined for objects of such
variability, but none are found.

Although not as deep at the WFPC2 observations, the (pre-servicing
mission) WF/PC images have better areal coverage of the error circle
and use two near-optimal filters, F336W (300 s, 900 s, 1200 s) and
F439W (100 s, 300 s, 400 s), those also used to identify several other
counterparts.  We concentrate here on the regions of the WF/PC images
which include the X-ray error circles.  In Fig. 2 we show a corner
section of one of the F439W exposures, with a cross placed at the X-ray
coordinates.  \decsec{5}{3} and \decsec{10}{2} error circles are
displayed as in Fig.  1.  All images have been cleaned of cosmic-ray
hits by an algorithm written by E.W.D.  The images were then carefully
inspected and any shortcomings of this algorithm were repaired
manually; stars which have been clearly hit by a cosmic-ray event were
excluded from the color-magnitude diagram.

Initial photometric zero points were taken from the PHOTFLAM keywords
in the headers.  However, another correction was applied to calibrate
the WF/PC $m_{439}$ magnitudes to the WFPC2 $m_{439}$ magnitudes.  This
correction was calculated using a set of isolated stars in common
between images from both cameras, and is primarily an aperture
correction, which is difficult to determine for WF/PC images.  This
same correction was applied to the $m_{336}$ magnitudes.

%==============================================================================
\section{DISCUSSION}

\subsection{Color-Magnitude Diagram}

Figure 3 shows a color-magnitude diagram derived from aperture
photometry of the WF/PC F336W and F439W observations.  Included are the
\squig 150 objects in PC6 which are detected in the F336W data.  Stars
within $r<\decsecmm{2}{5}$ from the cluster center are excluded as the
crowding there causes large errors in the photometry.  All objects in
this excluded region are $>18''$ from the X-ray position, and thus this
procedure does not affect the search for the optical counterpart.
Magnitudes are in the STMAG system, and formal $1\sigma$ error bars are
provided for each source.  Non-negligible flat-fielding imperfections
as well as the numerous low-level cosmic-ray hits will cause some
sources to be in error by more than the estimated uncertainties.

We overlay an isochrone from Bertelli et al. (1994) for comparison.
The 12~Gyr, [Fe/H$]=-0.7$ isochrone has been converted to the STMAG
system (conversion factors derived using the STSDAS {\it synphot}
package), corrected for $({\rm m-M})_0=14.85$, and reddened by
E($B-V)=0.10$.  The isochrone does not fit in detail but indicates that
the distance, reddening, and metallicity parameters yield an isochrone
which follows the observed points reasonably well, particularly given
the uncertainties in the filter transformations and absolute
calibration.  It can be clearly seen, however, that these data do not
even reach the main-sequence turnoff adequately.  The
$M_{B_0}$ scale incorporates the filter correction $(B-m_{439})=0.50$,
although the true correction should vary slightly with stellar color;
0.65 is appropriate for stars hotter than type F, and the correction
decreases to 0.40 for M0 stars.  We also plot open circles for the
distance- and reddening-adjusted magnitudes of the five known optical
counterparts of luminous globular cluster X-ray sources (Deutsch et al.
1998), if they were relocated to NGC\,6652.

A population of stars is seen blueward of the red giant branch in the
color-magnitude diagram of Fig. 3.  Each of these stars was examined
individually, and none are obviously due to cosmic-ray hits or other
defects in the data.  Most of these objects are, without question,
stars bluer than the red giant branch, although their nature is not
obvious.  These stars do fall in the region where one expects blue
straggler stars (BSS) based on the extension of the main sequence of
the isochrone, and as Ortolani et al.  (1994) made a special note
concerning the abundance of BSS in this cluster, we entertain the idea
that many of these stars might be BSS.

A detailed comparison between the blue stragglers discovered by
Ortolani et al. (1994) (obtained in electronic form from CDS) and blue
objects isolated here, shows that four of our blue objects correspond
directly to four BSS (OBB 772, 857, 858, 865) discovered by Ortolani et
al.  The other 5 BSS candidates from Ortolani et al. which fall in the
PC6 frame do not correspond to remarkable objects in our
color-magnitude diagram, and most may be inaccurately photometered
blends in the ground-based data.  The remaining dozen blue objects in
Fig. 3 are crowded by neighboring stars such that they would be very
difficult to isolate in ground-based images.  In addition, these
objects are more numerous near the center of the cluster than in other
parts of the frame and many, therefore, are likely to be cluster
members.

As the contamination by nonmembers is not negligible, we briefly
examine the expected contamination from such stars in the
color-magnitude diagram.  Although the {\it HST} fields are too close
to the cluster to study the contamination with these data, we estimate
from the contamination investigation in Fig. 5 of Ortolani et al.
(1994) that \squig 10 bulge stars $m_{439}<20$, and only a few bulge
stars $m_{439}<18$ are expected to appear in PC6.  We conclude that a
few of the blue stars in our diagram may be nonmembers, but most appear
to be cluster members and are likely to be blue stragglers.  In any
event, none of these blue stars are within the X-ray error circles, and
these objects are unlikely to be related to X1832--330.

\subsection{Optical Counterpart Candidates}

Within the partial coverage of the \decsec{5}{3} radius $1\sigma$ error
circle of the {\it ROSAT} HRI coordinates, there are no obvious
UV-excess candidates with $m_{439}<20.6$ ($M_{B_0}<5.9$), a limit
slightly stronger than the least luminous known globular cluster
optical counterpart, Star A in NGC\,1851 (Deutsch et al. 1996), when
all counterparts are corrected for distance and reddening.  However, it
is important to stress that the region beyond $4''$, while completely
covered by our AAT data, is not completely imaged by these {\it HST}
frames (Fig. 2); a faint optical counterpart mischievously located
$>4''$ due west of the nominal coordinates could not have been seen
with existing {\it HST} observations.

In the \asecbyasec{35}{35} region examined here, only one object stands
out as a candidate based on similar color to the other known
counterparts.  This faint UV-excess object, which we denote Star 49, is
the only unusual star in this region, and we therefore offer it as a
possible optical counterpart to the luminous globular cluster X-ray
source X1832--330.  From Fig. 3 it can be seen that Star 49 has color
similar to the other known counterparts, and a luminosity nearly equal
to Star A in NGC\,1851 (Deutsch et al. 1998).  For Star 49 object we
measure $m_{336}=19.69\pm0.06$, $m_{439}=20.2\pm0.2$, $(m_{336}-m_{439})=-0.5\pm0.2$, and apply
approximate filter corrections to estimate $B_0=20.4$, $(U-B)_0=-0.9$,
and $M_{B_0}=5.5$.

However, for Star 49 we measure a position in the GSC frame of
$\alpha(2000)={\rm18^h35^m}$\decsectim{44}{57},
$\delta(2000)={\rm-32^\circ59'}$\decsec{38}{3}, which is \decsec{11}{7}
away from the X-ray coordinates, at a distance greater even than the
estimated \decsec{10}{2} $2\sigma$ uncertainty.  About 2\% of {\it
ROSAT} HRI aspect solutions seem to be in error by more than $12''$, so
this object should not be discounted, although it must be treated with
caution.

Star 49 itself is near the detection threshold in both filters.  The
object does, however, appear to be weakly detected in four separate
exposures, two F439W frames and two F336W frames (each filter pair at a
different spatial offset).  It is too faint to be expected to be
detected in the other two shorter-exposure {\it HST} images, and indeed
does not appear.  In Fig. 4 we show coadded F336W and F439 frames of
Star 49 and the surrounding \asecbyasec{8}{8} region.  The existing
WFPC2 observations are not helpful for further studying this object, as
it falls just outside the WFPC2 field of view.

%==============================================================================
\section{CONCLUSION}

We have presented a search of the optical counterpart for X1832--330,
the luminous globular cluster X-ray source in NGC\,6652.  Using the GSC
reference frame, we determine the optical position of the {\it ROSAT}
X-ray coordinates on a CCD image.  U and B ground-based images from the
AAT allow us to set a limit $M_{B_0}\squiggeqmm3.5$ for the counterpart
at the time of those observations, provided the color is $(U-B)_0=-1$.
Archival {\it HST} WFPC2 exposures which subtend most but not all of the error
circle allow us to infer $M_{B_0}>5.2$ for the counterpart if in that
region, again provided at the time of observation the source is
UV-excess $(m_{218}-m_{439})_0=-1.4$, like the counterpart in
NGC\,6441.  Archival WF/PC observations allow a more sensitive search;
within the \squig90\% of the \decsec{5}{3} radius $1\sigma$ error
circle about the X-ray coordinates contained in the WF/PC images, we
detect no objects at $m_{439}<20.6$ ($M_{B_0}<5.9$) with colors
compatible with the other known optical counterparts in globular
clusters.  The region outside radius $4''$ is not completely imaged by
these data, and therefore a faint UV-excess counterpart could have been
missed with these {\it HST} observations.

We do weakly detect a faint UV-excess object \decsec{11}{7} from the
{\it ROSAT} coordinates.  This is a $2.3\sigma$ deviation from the
X-ray coordinates, and thus this object certainly should not be
completely ruled out based on its position.  If it is indeed the correct
identification, this object provides another example of the extremely
underluminous optical counterpart seen in NGC\,1851.  We measure for
Star 49 $m_{439}=20.2\pm0.2$, $(m_{336}-m_{439})=-0.5\pm0.2$, and
estimate $B_0=20.4$, $(U-B)_0=-0.9$, and $M_{B_0}=5.5$.

Should the X-ray coordinates prove to be accurate to better than $4''$,
the likely conclusion is that the true optical counterpart, not yet
identified, is the intrinsically faintest cluster source yet known, at
least at the time of these observations, and Star 49 may be yet another
example of a faint UV-excess cluster object of unknown nature (see,
e.g., Deutsch et al. 1996, 1998).  Another possibility is that the
optical light from the system was not dominated by the hot accretion
disk at the time of these observations, but rather by the secondary,
thereby rendering the object's color unremarkable in our
color-magnitude diagram; however, such behavior has not been observed
for the other identified cluster sources.  Clearly, deep F336W, F439W
WFPC2 observations placing the X-ray coordinates in the center of the
PC chip are desirable to search this field more thoroughly, and a more
accurate X-ray position would reduce the number of objects which must be
considered as possible candidates.

Identification of optical counterparts in the remaining clusters with
luminous X-ray sources will be difficult using current techniques, as
is seen in Fig. 5.  Here we have plotted the $m_{336}$ apparent
magnitudes, either observed or predicted, of the optical counterparts
of luminous X-ray sources in the cores of globular clusters.  The
objects are ordered by right ascension, so the units on the abscissa
are of no significance.  Clusters with boxed names are those with
optical counterparts already identified (or tentatively suggested in
this work); all but one required {\it HST} observations.  The dashed
vertical lines denote the 5~mag range of the luminosity dispersion
implied by the current complement of identifications, adjusted for the
distance and reddening of each cluster (cluster parameters principally
from the compilation of Djorgovski 1993).  The filled squares are the
observed $m_{336}$ magnitudes derived from our {\it HST} photometry
(Deutsch et al. 1998); the positions of the squares within the dashed
lines indicate the luminosities compared with the luminosity range of
all the known sources (e.g., the object in NGC\,1851 appears at the
bottom of the dashed line as it is the least luminous one).  For the
six sources with no current identification, the six open diamonds
indicate where each of the known identifications would fall if
relocated to the target cluster, again with appropriate distance
modulus and reddening.  The horizontal line at $m_{336}=23$ denotes the
approximate flux reached by a typical short WFPC2 program, i.e., 10\%
photometric precision in a two-orbit multicolor exposure series in a
moderately crowded field.

Of the ``easy" clusters, i.e. those for which the optical counterpart
can be expected to be detected in one or two {\it HST} orbits,
NGC\,6652 is the last for which a candidate has been put forth.  Due to
the considerable foreground extinction, the remaining unidentified
luminous cluster X-ray sources will be difficult to identify with current
techniques at the UV and blue wavelengths where these sources have in
the past been studied.  A possible exception is NGC~6440, which
could be adequately studied in a few orbits.  For the remaining
clusters, the amount of time required to detect a low-luminosity
counterpart in an F336W frame with the WFPC2 becomes prohibitive,
although detection of counterparts at the high end of the luminosity
range is feasible.  Variability searches in the infrared may afford a
way to identify and study the remaining, heavily-reddened sources.
Future searches will also be facilitated when arcsecond-accuracy X-ray
coordinates become available via AXAF observations, thereby drastically
reducing the number of optical objects which must be considered.
Nonetheless, there still remains much to be learned from the current
crop of optical counterparts, just discovered in the last few years.

%==============================================================================
\acknowledgments

We thank Stephanie Wachter for providing ground-based CCD images of
NGC\,6652, Joss Bland-Hawthorn for assisting with the TTF observations,
and Wolfgang Voges for helping us quantify {\it ROSAT} HRI aspect
uncertainties.  Support for this work was provided by NASA Grant
NAG5-1630.

%==============================================================================

\clearpage

\begin{figure}
%\plotone{../ctio/figure1.ps}
\plotone{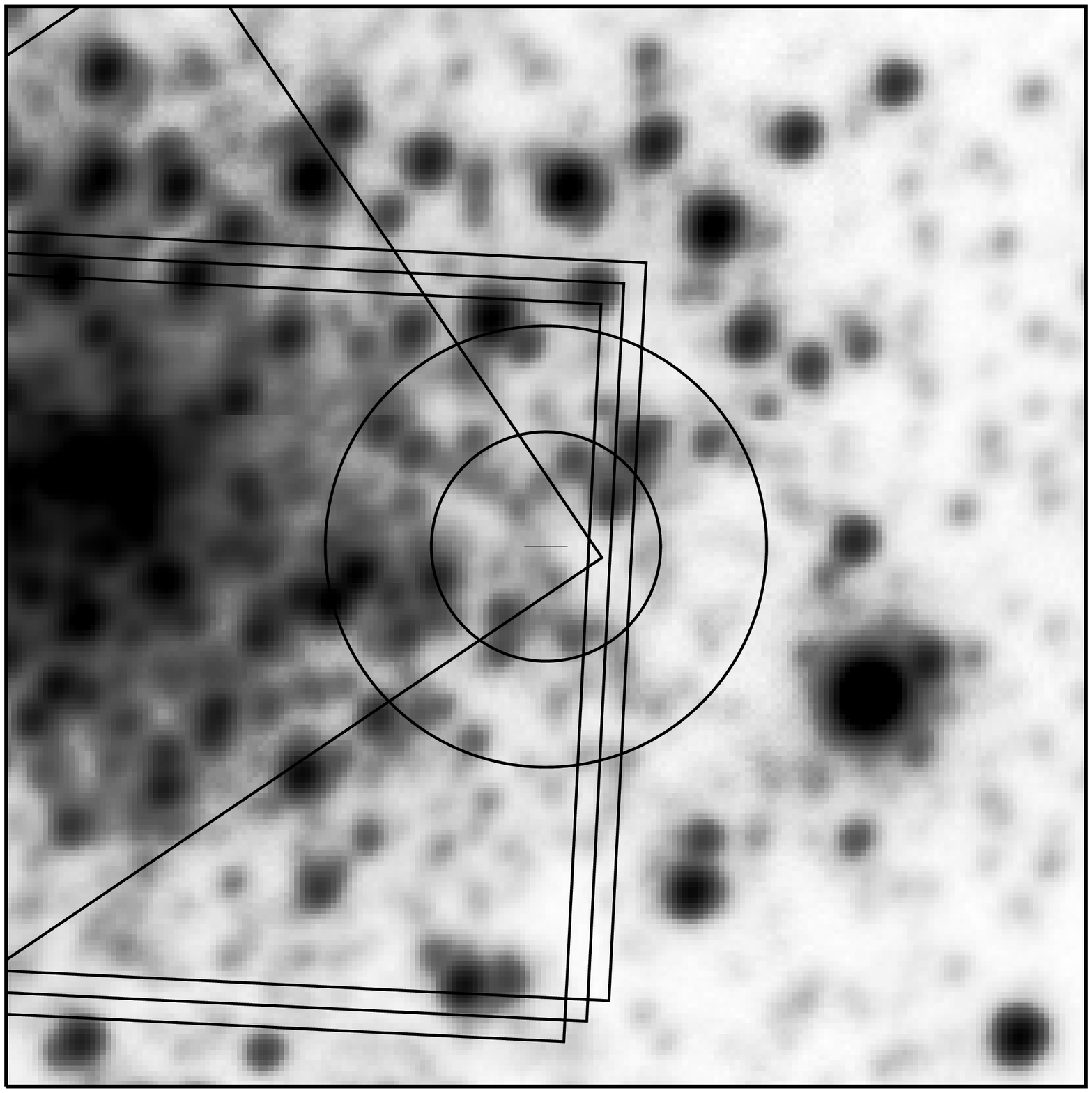}
\caption{\asecbyasec{50}{50} ground-based B-band CCD image centered
(cross) on the {\it ROSAT} HRI coordinates for the luminous globular
cluster X-ray source X1832--330.  North is up and East is left.
Overlaid are \decsec{5}{3} $1\sigma$ and \decsec{10}{2} $2\sigma$
radius X-ray error circles.  The image was obtained in \decsec{1}{3}
seeing conditions, and shows that the field is quite crowded
from the ground.  Also overlaid are the multiple archival {\it HST}
exposure fields of view.  The WF/PC observations were obtained at three
different offsets and exposure times for each filter, while the WFPC2
observations were all at the same pointing.  Unfortunately, the X-ray
error circles extend beyond the boundaries of the {\it HST} frames.}
\end{figure}

\begin{figure}
%\plotone{../wfpc1/figure2.ps}
\plotone{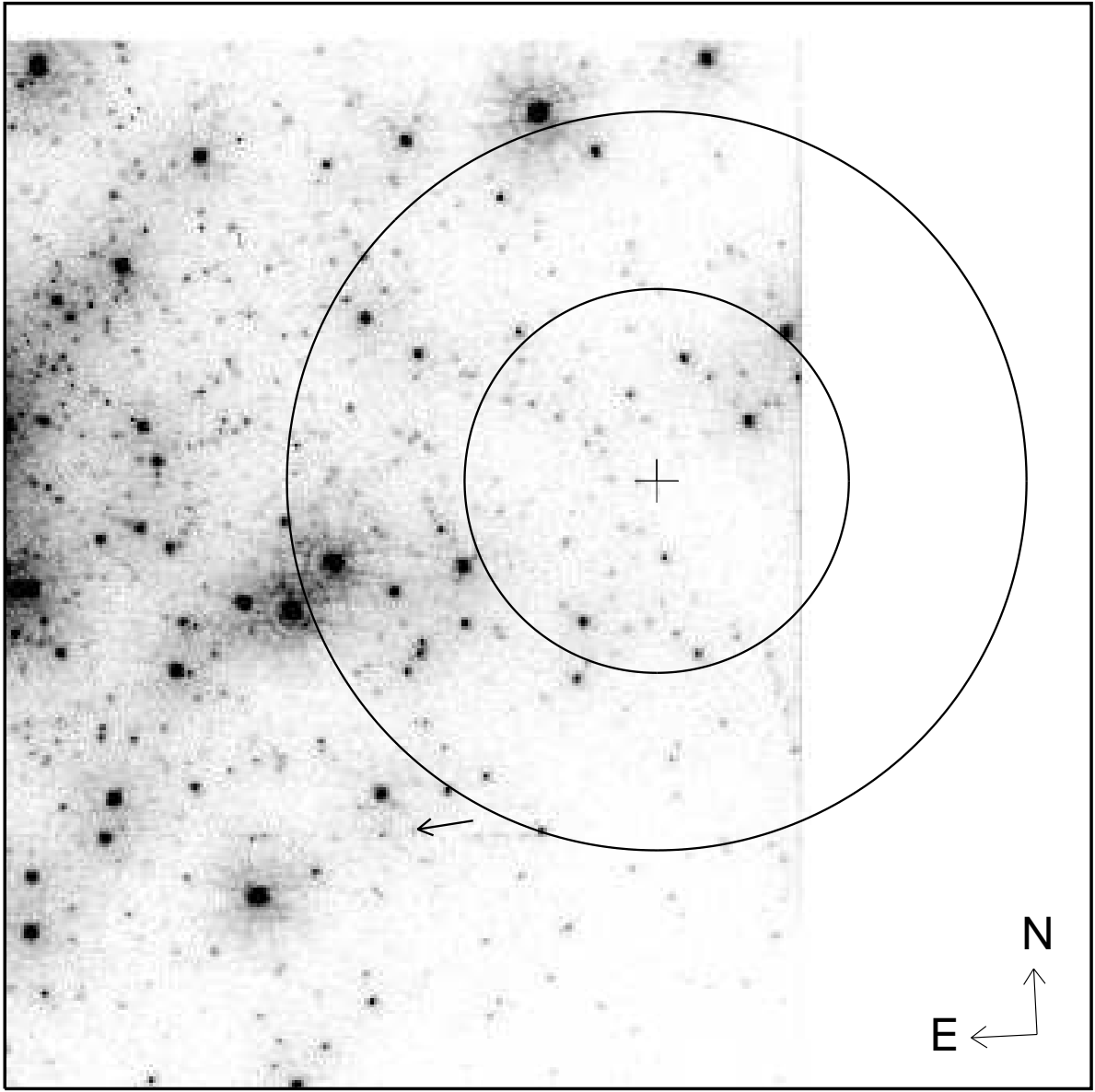}
\caption{\asecbyasec{22}{29} corner section of the 300 s F439W ($B$)
WF/PC image, which has been cleaned of cosmic-ray hits.  Our optical position
for the X-ray coordinates is marked with a cross.  \decsec{5}{3}
$1\sigma$ and \decsec{10}{2} $2\sigma$ error circles are displayed as
in Fig. 1.  The field is not severely crowded by {\it HST} standards,
although there are still \squig40 objects within the $1\sigma$ error
circle.  A faint object, discussed in the text as a possible counterpart,
is marked with an arrow.}
\end{figure}

\begin{figure}
%\plotone{../wfpc1/figure3.ps}
\plotone{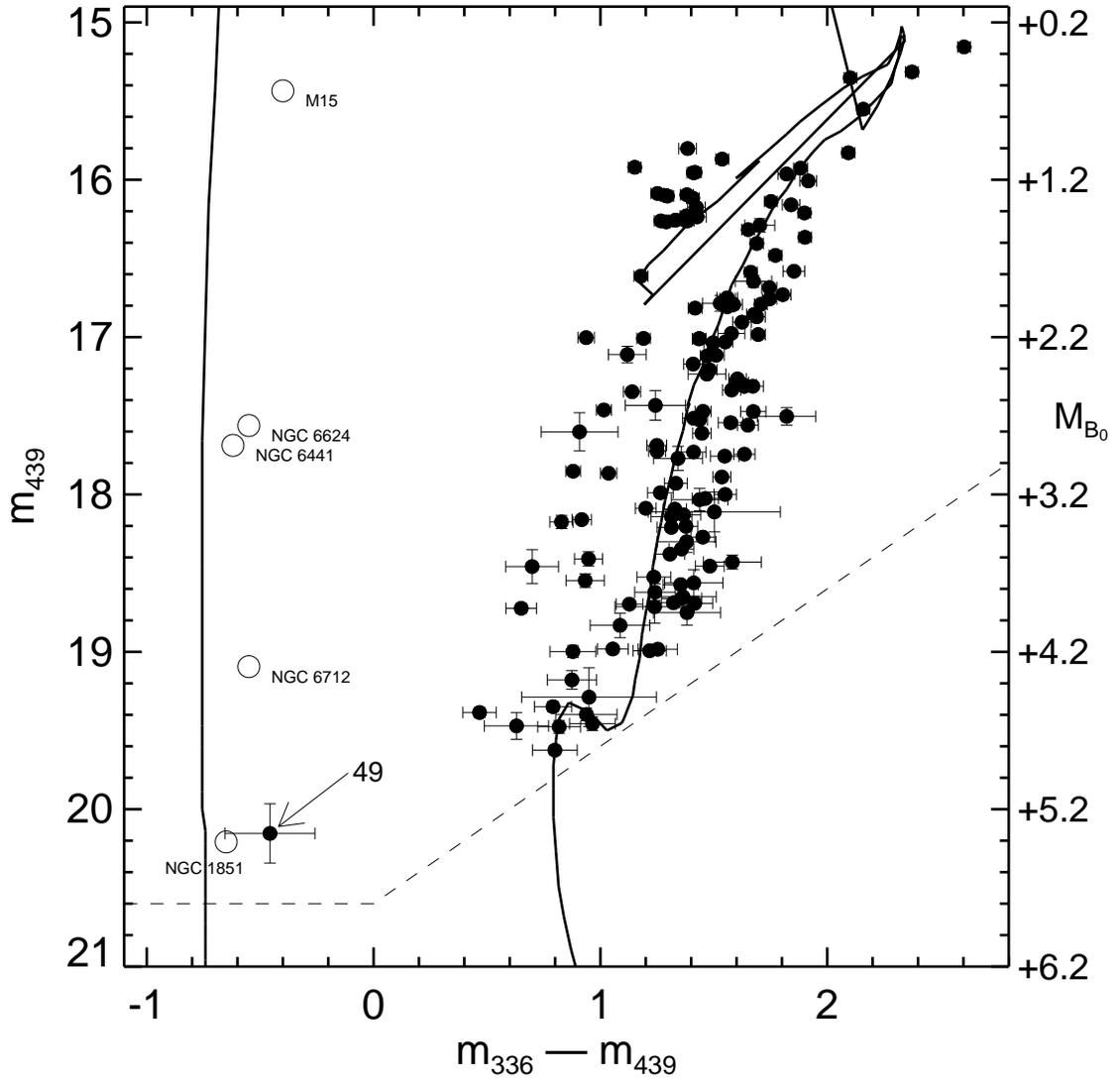}
\caption{Color-magnitude diagram derived from the WF/PC F336W and F439W
observations for the \squig 150 objects in PC6 which are detected in both
passbands.  Magnitudes are in the STMAG system, and formal $1\sigma$
error bars are provided for each object, although a few objects will
have greater error due to low-level cosmic-ray hits which could not be
identified and flagged.  Solid line:  a 12 Gyr, [Fe/H$]=-0.7$ isochrone
from Bertelli et al. (1994), added for comparison.  Dashed line:
approximate detection threshold for these data.  The distance- and
reddening-adjusted locations for the five known optical counterparts of
luminous globular cluster X-ray sources (Deutsch et al. 1998) are
plotted as open circles.  Only one object near the X-ray source, which
we denote Star 49, has color and magnitude similar to other known optical
counterparts.}
\end{figure}

\begin{figure}
%\plotone{../wfpc1/figure4.ps}
\plotone{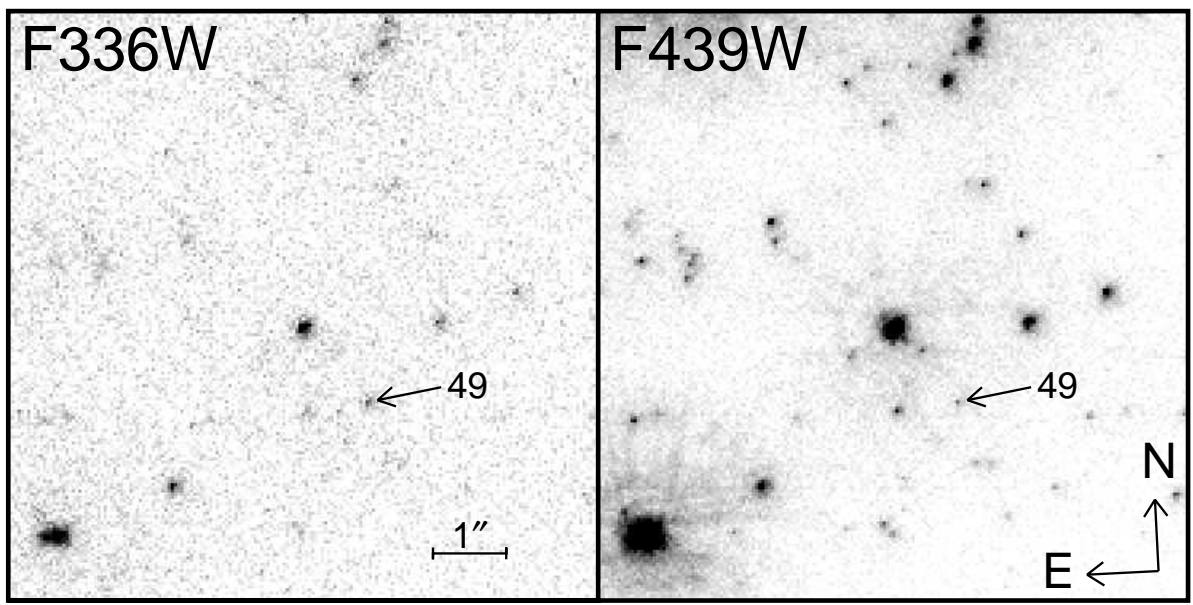}
\caption{\asecbyasec{8}{8} sections of the F336W ($U$) and F439W ($B$)
images of the region near Star 49, which we select as a possible
optical counterpart due to significant UV excess, similar to other
known cluster optical counterparts.  These {\it HST} WF/PC data were
obtained on 1992 October 10.  Each frame is the sum of the two
longest images for each filter.}
\end{figure}

\begin{figure}
%\plotone{../figures/figure5.ps}
\plotone{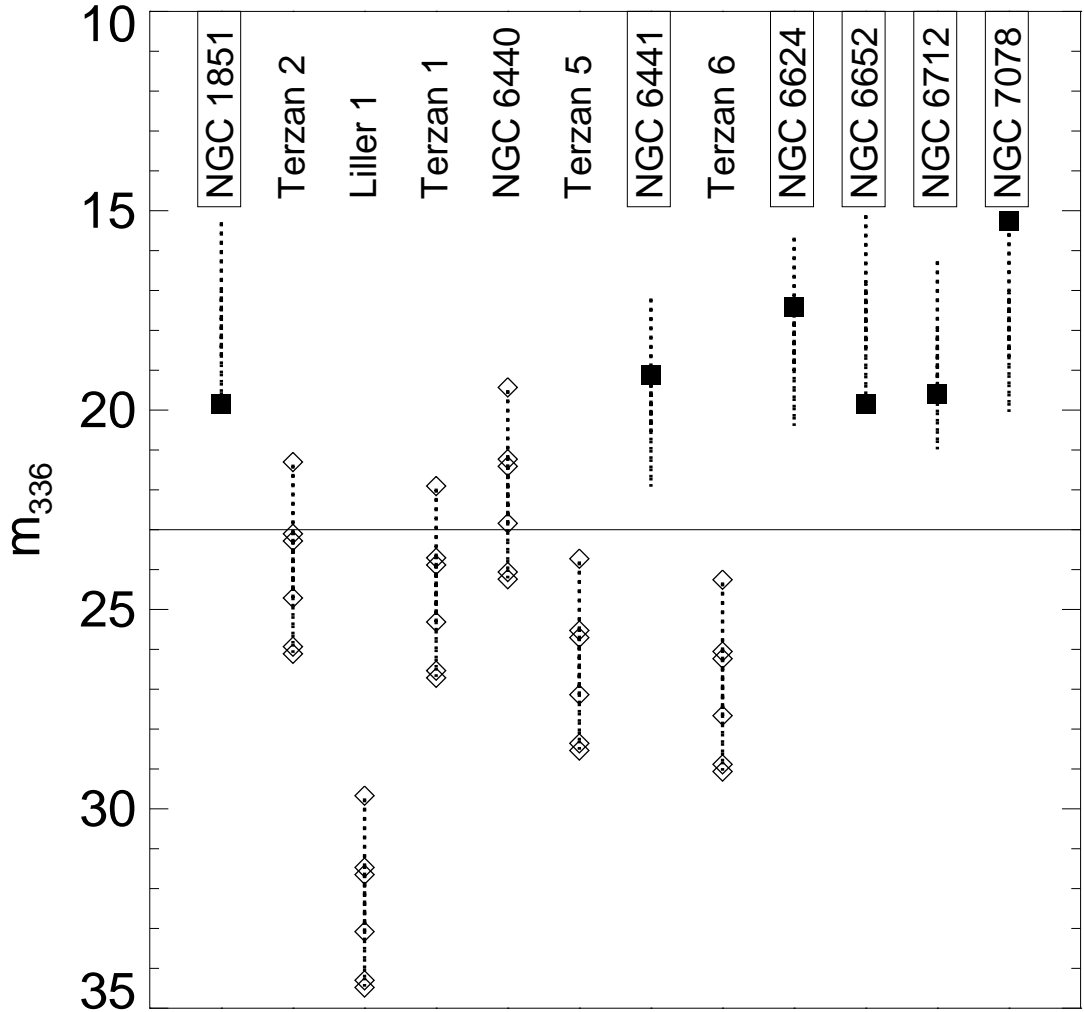}
\caption{Apparent $m_{336}$ magnitudes for the six confirmed or
suggested optical counterparts to luminous globular cluster X-ray
sources (filled squares), full range of optical luminosities of the
ensemble of the current candidates (dashed vertical lines), and
predicted magnitudes for each of the known identifications if relocated
to other target clusters with appropriate distance modulus and
reddening (open diamonds).  The horizontal line at $m_{336}=23$ denotes
an estimate of the S/N=10 limit in a sequence of $4\times700$~s WFPC2
F336W exposures in moderate crowding conditions, i.e. part of a typical
short WFPC2 program.  Of the ``easy" clusters, i.e. those for which the
luminosity range is almost entirely above the short {\it HST} program
detection limit, NGC\,6652 is the last for which a candidate has been
suggested.}
\end{figure}

\end{document}